\documentclass[aip,reprint]{revtex4-1}
\usepackage{graphicx}
\usepackage{bm, physics}
\usepackage{float}
\usepackage{mathtools}
\usepackage{hyperref}
\usepackage{color}

\let\svthefootnote\thefootnote
\newcommand\freefootnote[1]{%
  \let\thefootnote\relax%
  \footnotetext{#1}%
  \let\thefootnote\svthefootnote%
}

\begin{document}

\title{Shift of quantum critical point of discrete time crystal on a noisy quantum simulator}
\author{Yuta Hirasaki}
\email{yutah2@illinois.edu}
 \affiliation{ 
Department of Applied Physics, The University of Tokyo, Tokyo 113-8656, Japan.
}%
 \affiliation{ 
Department of Physics, University of Illinois at Urbana-Champaign, Urbana, Illinois 61801, USA.
}%

\author{Toshinari Itoko}
\email{itoko@jp.ibm.com}

\author{Naoki Kanazawa}
\affiliation{%
IBM Quantum, IBM Research--Tokyo, 19-21 Nihonbashi Hakozaki-cho, Chuo-ku, Tokyo, 103-8510, Japan.
}

\author{Eiji Saitoh}
\affiliation{ 
Department of Applied Physics, The University of Tokyo, Tokyo 113-8656, Japan.
}%
\affiliation{Institute for AI and Beyond, The University of Tokyo, Tokyo 113-8656, Japan.}
\affiliation{WPI Advanced Institute for Materials Research, Tohoku University, Sendai 980-8577, Japan.}
\affiliation{RIKEN Center for Emergent Matter Science (CEMS), Wako 351-0198, Japan.}

\date{\today}

\begin{abstract}
Recent advances in quantum technology have enabled the simulation of quantum many-body systems on real quantum devices. However, such quantum simulators are inherently subject to decoherence, and its impact on system dynamics—particularly near quantum phase transitions—remains insufficiently understood. In this work, we experimentally investigate how decoherence in quantum devices affects the dynamics of quantum time crystals, using a 156-qubit IBM Quantum system. We find that decoherence shifts the location of critical behavior associated with the phase transition, suggesting that noisy simulations can lead to inaccurate identification of phase boundaries. Our results underscore the importance of understanding and mitigating decoherence to reliably simulate quantum many-body systems on near-term quantum hardware.
\end{abstract}


\maketitle 

Recent advances in quantum technology have enabled highly accurate control of quantum degrees of freedom, with state-of-the-art quantum computers now offering access to hundreds of qubits. Quantum simulators have emerged as a powerful platform for exploring complex quantum many-body systems. A growing body of research has demonstrated the potential of quantum simulators to uncover novel physical phenomena~\cite{bernien2017probing, shtanko2025uncovering, kim2023evidence, doi:10.1126/sciadv.abm7652, mi2022time, doi:10.1126/science.abq5769, PhysRevX.15.011055, shinjo2024unveiling, PhysRevX.15.011055, doi:10.1126/science.abi8794}. 
Notably, previously unexplored quantum phase transitions have been observed in such simulations~\cite{bernien2017probing}, 
Additionally, the dynamics of spin systems have been studied across a range of configurations, revealing rich behaviors such as entanglement growth and thermalization~\cite{kim2023evidence, doi:10.1126/sciadv.abm7652}. These studies have significantly deepened our understanding of quantum critical phenomena and many-body dynamics by identifying quantum phase transitions through singularities in experimentally measurable quantities.

Current quantum computers inevitably face limitations such as noise, decoherence, and gate operation errors. To address these challenges and enable simulations on noisy intermediate-scale quantum devices, a variety of techniques have been proposed, including quantum error suppression and quantum error mitigation~\cite{RevModPhys.95.045005, PhysRevX.8.031027, PhysRevX.11.031057, PhysRevX.7.021050, PhysRevLett.119.180509}. In particular, quantum error mitigation refers to a set of classical post-processing methods that reduce the effects of errors in quantum computations. These techniques have been shown to improve the accuracy of results obtained from noisy quantum devices, making it possible to obtain more reliable outcomes when applying quantum algorithms to complex problems. Several studies have applied quantum error mitigation to explore quantum many-body systems~\cite{shtanko2025uncovering, kim2023evidence, doi:10.1126/sciadv.abm7652}, enabling more precise estimation of observable expectation values. 
However, the impact of decoherence on critical quantum phenomena, such as phase transitions, remains insufficiently explored.

In this study, we investigate the impact of decoherence on quantum many-body dynamics. As a paradigmatic example of quantum matter realized on a quantum simulator, we focus on a one-dimensional discrete time crystal (DTC). To model decoherence, we simulate Pauli noise by randomly inserting Pauli gates into the time-evolution quantum circuit. We analyze how this noise influences the system’s dynamics, with particular attention to its effects on the DTC order parameter and the associated phase transition.


The discrete time crystal is a prototypical non-equilibrium phase of matter that has been extensively studied using quantum simulators~\cite{zhang2017observation, frey2022realization, choi2017observation, RevModPhys.95.031001}. While the existence of DTCs in equilibrium settings is prohibited~\cite{PhysRevLett.114.251603}, their realization in non-equilibrium regimes has opened new avenues for exploring quantum dynamics~\cite{PhysRevLett.118.030401}. Here, we analyze the model introduced in Ref.~\cite{frey2022realization} and simulate decoherence by randomly inserting Pauli $Z$ gates after each Floquet unitary time evolution, as described in detail below.

\begin{figure}
    \centering
    \includegraphics[width=1.0\linewidth]{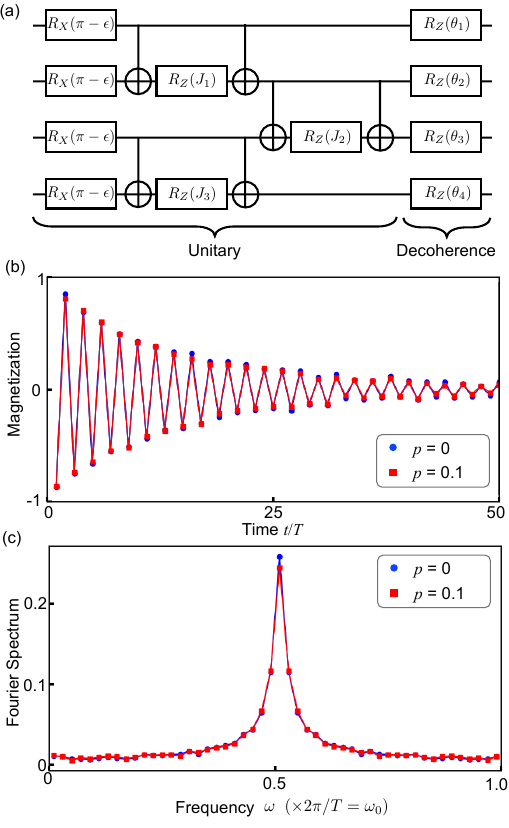}
    \caption{
    The experimental setup. (a) Quantum circuits employed in our experiments to simulate phase decoherence. The $Z$ gates are inserted randomly after the Floquet unitary to simulate the phase decoherence. (b) Real time oscillation of the magnetization. The result with (without) the decoherence $p = 0 (0.1)$ is depicted with the blue (red) dots. (c) The Fourier spectrum of the magnetization oscillation. The peak at the frequency $\omega = 0.5 \times\omega_0$ is the order parameter of the discrete time crystal.}
    \label{fig2}
\end{figure}

We consider a one-dimensional spin chain with the periodic Hamiltonian $H(t + T) = H(t)$
\begin{align}
H(t) = 
\begin{cases}
 -\frac{\pi}{2}(1-\epsilon)\sum_q X_q& 0\leq t< 1\\
 -\sum_q J_qZ_qZ_{q+1} & 1\leq t < 2 = T
\end{cases},
\end{align} 
where $q$ denotes the qubit index, $\epsilon$ is a small perturbation parameter that results in imperfect spin flip, and $J_q$ is a random interaction strength that gives rise to the many-body localization. The time evolution operator in one period is given by 
\begin{align} 
    U(T) = \prod_{q}\exp\left(i J_qZ_{q}Z_{q + 1}\right)\prod_{q}\exp\left[i\frac{\pi}{2}(1-\epsilon)X_q\right],
\end{align} 
and this unitary operator can be expressed using $R_X$ gate, controlled NOT gate, and $R_Z$ gate as shown in Fig.~\ref{fig2}(a). 


We now describe how we simulate the effects of decoherence. Quantum circuits are subject to various sources of decoherence and noise; however, it has been shown that such noise can be effectively modeled as stochastic Pauli errors~\cite{PhysRevA.94.052325}. These Pauli errors can be simulated on quantum devices by randomly inserting Pauli gates into the circuit. This method has been widely employed in quantum error mitigation to construct effective Pauli channels~\cite{kim2023evidence, van2023probabilistic}.

Following this approach, we simulate the effect of phase decoherence by randomly inserting single-qubit $Z$ gates after the Floquet unitary.
As shown in Fig.~\ref{fig2}(a),
we apply $R_Z$ gates with rotation angles randomly chosen to be $\pi$ (with probability $p$) or $0$ (with probability $1 - p$), corresponding to the Pauli-$Z$ gate and the identity, respectively.
By averaging the measurement outcomes over an ensemble of circuits with randomly sampled parameters $\{\theta_q\}_q$, we effectively simulate the quantum channel. 
\begin{align}
    \mathcal{E} = \bigcirc_q\mathcal{E}_q, \quad \mathcal{E}_q(\rho) = (1-p)\rho + pZ_q\rho Z_q.
\end{align}

We initialize the system in 
the computational basis state $\ket{0}^{\otimes n}$ where $n$ is the number of qubits and measure the averaged magnetization $\sum_q \langle Z_q \rangle / N_q$
at each time step $t = T, 2T, \dots$, where $N_q$ denotes the total number of qubits. We analyze the time evolution of the magnetization to characterize the system’s dynamical behavior. In the time-crystalline phase, the magnetization is expected to exhibit oscillations with a period of $2T$, and the associated decay time grows exponentially with system size.

We define the order parameter $h$ as the amplitude of the Fourier spectrum at frequency $\omega = \omega_0/2 = \pi/T$, where $\omega_0 = 2\pi/T$ is the driving frequency. This frequency component corresponds to the characteristic subharmonic response of the discrete time crystal. The phase transition point is identified by analyzing fluctuations in the order parameter $h$ across different realizations of disorder, characterized by varying random interaction strengths $\{J_i\}_i$ .

We analyze a 70-qubit system evolved over 20 time steps, using 100 values of the coupling constant $J$ sampled uniformly from the range $[\frac{\pi}{4}, \frac{3\pi}{4}]$ , along with 99 different random configurations of $Z$ gates.
In total, we generate 198{,}000 quantum circuits ($20 \times 99 \times 100$), each of which is sampled 10 times on the IBM Quantum Heron processor, \texttt{ibm\_aachen}.

We measure the real-time magnetization oscillation and its Fourier spectrum at decoherence rates $p = 0, 0.1$, and present the results in Fig.~\ref{fig2}(b)(c). In Fig.~\ref{fig2}(b), the magnetization oscillation 
is depicted as a function of the time. 
The corresponding Fourier spectrum is shown in Fig.~\ref{fig2}(c). The spectral amplitude at frequency $\omega = \omega_0/2$ corresponds to an oscillation with period $2T$. The height $h$ of this spectral peak serves as the order parameter for discrete time crystals~\cite{PhysRevLett.118.030401}. We observe that the order parameter $h$ decreases with increasing decoherence rate $p$, suggesting that the time-crystalline order is fragile against noise. We note that this is consistent with the theoretical prediction\cite{PhysRevB.95.195135}.

\begin{figure}
    \centering
    \includegraphics[width=1.0\linewidth]{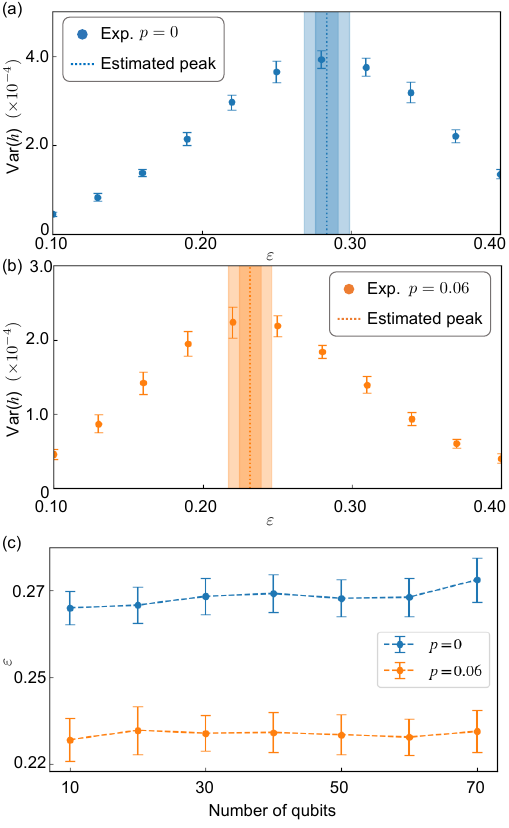}
    \caption{The shift in the criticality. (a) The variance of the order parameter as a function of the parameter $\varepsilon$ without the decoherence $p = 0$.
    The vertical dashed line shows the estimated peak points using smoothing cubic spline fitting, and the neighboring bands show the 1-sigma and 2-sigma intervals.
    (b) The variance of the order parameter with the decoherence $p = 0.06$.
    (c) The position of the peak height varying the number of qubits. The blue (orange) dots show that without (with) decoherence $p = 0 (0.06)$.
    The experiment was conducted using 70 qubits in an IBM Quantum system, and we reproduced the results in (a) and (b). The result in (c) was reproduced using only a subset of qubits. 
    }
    \label{fig3}
\end{figure}

We next investigate the critical behavior of discrete time crystals near the quantum phase transition point. Previous studies have shown that the variance of the order parameter exhibits critical fluctuations at the transition, when the random coupling constants are sampled from a specific distribution~\cite{frey2022realization, PhysRevLett.118.030401}. In our study, we measure the variance of the order parameter under two different phase decoherence strengths, $p = 0$ and $p = 0.06$, and present the results in Fig.~\ref{fig3}.

For each of the 99 random $Z$-gate configurations, the resulting data are divided into 20 batches. Within each batch, we compute the variance of the order parameter for each value of $\epsilon$, and estimate the peak position by fitting a smoothing cubic spline curve. The mean of the 20 estimated peak positions, along with the corresponding 1-sigma and 2-sigma confidence intervals, is shown in Fig.~\ref{fig3}.

Figure~\ref{fig3}(a) shows the variance of the order parameter without decoherence, $p = 0$. The variance of $h$ peaks around $\epsilon \approx 0.284.$ Figure~\ref{fig3}(b) shows the variance for decoherence strength $p = 0.06$, where the peak lies at $\epsilon \approx 0.231.$ There is a statistically significant shift in the peak position due to the introduction of decoherence. This shift indicates that decoherence alters the location of the critical fluctuation associated with the quantum phase transition.

We investigate finite-size effects by varying the system size from 10 to 70 qubits, post-selecting subsets of qubits from the full system. 
The estimated peak positions of the order parameter variance are shown in Fig.~\ref{fig3}(c). The blue dots represent the peak positions without decoherence ($p = 0$), while the orange dots correspond to those with phase decoherence ($p = 0.06$). Error bars indicate the standard deviation of the measurements across different random configurations. The results show that the peak position remains essentially independent of the system size. Therefore, the observed shift due to decoherence is expected to persist in the thermodynamic limit.

To validate the experimental results, we performed classical numerical simulations by varying the parameter $\epsilon$ and the decoherence strength $p$, and analyzed the behavior of the variance of the order parameter $h$. The simulations were carried out by tracking the full density matrix for a system size of $N = 12$. Here we set the open boundary condition. 

In the numerical simulation, we prepared 1000 samples of random coupling configurations $\{J_i\}_i$, sampled from the range $[\pi/4, 3\pi/4]$,
and computed the variance of the order parameter for each setting. The parameter $\epsilon$ was swept from 0 to 0.5 in 30 evenly spaced intervals, and the decoherence strength $p$ was varied from 0 to 0.1 in 30 increments. We kept track of the full density matrix classically over 50 time steps.

\begin{figure}
    \centering
    \includegraphics[width=1.0\linewidth]{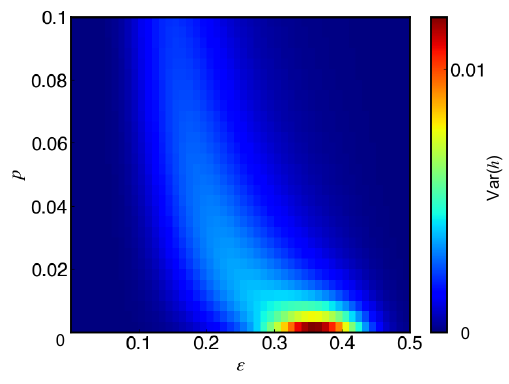}
    \caption{The variance of the order parameter as a function of the unitary perturbation $\epsilon$ and the decoherence $p$,
    obtained by the numerical simulation. We calculated the order parameter by tracking the full density matrix with the qubit size $12$. The number of coupling constant configuration prepared is $1, 000$. The order parameter is computed from the magnetization oscillations over 50 time steps. The coupling constants ${J_i}_i$ are independently sampled from a uniform distribution over the interval $[\pi/4, 3\pi/4]$.
    }
    \label{fig4}
\end{figure}

Figure~\ref{fig4} presents the variance of the order parameter as a heatmap obtained by the numerical simulation. The horizontal axis corresponds to the unitary perturbation parameter $\epsilon$, while the vertical axis represents the decoherence strength $p$. The variance of the order parameter is depicted as a color gradient. In the absence of decoherence ($p = 0$)), the variance reaches a maximum around $ \epsilon \approx 0.35$ , as indicated by the red-colored region in the lower right portion of the figure. As the decoherence strength $p$ increases, we observe both a gradual suppression and a shift in the location of the peak variance. The critical fluctuation of the order parameter diminishes with increasing decoherence, as shown by the transition in the color gradient from warmer to cooler shades. In addition, the peak position shifts to smaller values of $\epsilon$, as illustrated by the curvature of the cooler-shaded region. This behavior is consistent with the experimental results shown in Fig.~\ref{fig3}. It is worth noting that the classical numerical simulations performed here are free from decoherence when $p = 0$, whereas the quantum computer experiments in Fig.~\ref{fig3} inevitably experience intrinsic noise, even at $p = 0$.

In this study, we have investigated the effect of decoherence on the critical behavior of quantum many-body systems. We simulated the phase decoherence by randomly inserting Pauli $Z$ gates into quantum circuits. Our results show that decoherence not only drives the system toward a maximally mixed state, but also alters the system's critical properties—shifting the estimated location of critical points.

Our findings suggest that the effects of decoherence may be mitigated using a strategy analogous to probabilistic error amplification \cite{kim2023evidence}. Specifically, as shown in Fig.~\ref{fig4}, the shift in the peak of the critical fluctuation can be characterized as a function of the decoherence strength $p$. This decoherence strength can be artificially amplified by randomly inserting Pauli gates, as demonstrated in this work and in previous studies on probabilistic error amplification and zero-noise extrapolation. A detailed analysis of the curve in Fig.~\ref{fig4}, along with a precise characterization of the Pauli error model on the experimental hardware, will be essential for implementing such an approach effectively. We leave this investigation for future research.

A theoretical understanding of the observed shift in criticality is desirable. In particular, the effect of decoherence on quantum critical phenomena could be further explored using alternative observables~\cite{PhysRevLett.121.016801, PhysRevB.107.L020202, kobayashi2025quantum, 10.21468/SciPostPhys.18.6.198}. From an experimental perspective, it would also be interesting to investigate how time-dependent noise influences critical behavior~\cite{burnett2019decoherence, PhysRevLett.121.090502, PhysRevLett.123.190502, 10.1063/5.0166739, 10.1063/5.0226517, etxezarreta2021time, stehli2020coherent}. These directions offer promising avenues for future work to deepen our understanding of the interplay between decoherence and quantum phase transitions.

\section{Acknowledgment}
The authors thank Oles Shtanko and Kunal Sharma for insightful comments. 
Y. H. thanks Shunsuke Daimon and Naoto Yokoi for fruitful discussions. This work was partially supported by CREST (No. JPMJCR20C1, No. JPMJCR20T2) from JST, Japan; Grant-in-Aid for Transformative Research Areas (No. JP22H05114) from JSPS KAKENHI, Japan. This work was also supported by Sumitomo Chemical. Y.H. was partially supported by Murata Overseas Scholarship and KDDI foundation Overseas Scholarship. This work is partially supported by the IBM-Utokyo laboratory. We acknowledge the use of IBM Quantum services for this work. The views expressed are those of the authors, and do not reflect the official policy or position of IBM or the IBM Quantum team.

\section*{AUTHOR DECLARATIONS}%
\subsection*{Conflict of Interest}
The authors have no conflicts to disclose.

\subsection*{Author Contributions}
{\bf{Y. Hirasaki}}: Conceptualization (equal); Formal analysis (lead); Investigation (equal); Methodology (equal); Software(equal); Validation (equal); Writing – original draft (lead).
{\bf{T. Itoko}}: Investigation (equal); Methodology (equal); Software(equal); Supervision (supporting); Validation (equal); Writing – review \& editing (supporting).
{\bf{N. Kanazawa}}: Project administration (supporting); Software(supporting); Supervision (supporting); Writing – review \& editing (supporting).
{\bf{E. Saitoh}}: Conceptualization (equal); Funding acquisition (lead); Project administration (lead); Supervision (lead); Validation (equal); Writing – review \& editing (supporting).

\section*{Data Availability Statement}
The data that support the findings of this study are available from the corresponding author upon reasonable request.

\bibliography{manuscript}

\begin{thebibliography}{34}%
\makeatletter
\providecommand \@ifxundefined [1]{%
 \@ifx{#1\undefined}
}%
\providecommand \@ifnum [1]{%
 \ifnum #1\expandafter \@firstoftwo
 \else \expandafter \@secondoftwo
 \fi
}%
\providecommand \@ifx [1]{%
 \ifx #1\expandafter \@firstoftwo
 \else \expandafter \@secondoftwo
 \fi
}%
\providecommand \natexlab [1]{#1}%
\providecommand \enquote  [1]{``#1''}%
\providecommand \bibnamefont  [1]{#1}%
\providecommand \bibfnamefont [1]{#1}%
\providecommand \citenamefont [1]{#1}%
\providecommand \href@noop [0]{\@secondoftwo}%
\providecommand \href [0]{\begingroup \@sanitize@url \@href}%
\providecommand \@href[1]{\@@startlink{#1}\@@href}%
\providecommand \@@href[1]{\endgroup#1\@@endlink}%
\providecommand \@sanitize@url [0]{\catcode `\\12\catcode `\$12\catcode `\&12\catcode `\#12\catcode `\^12\catcode `\_12\catcode `\%12\relax}%
\providecommand \@@startlink[1]{}%
\providecommand \@@endlink[0]{}%
\providecommand \url  [0]{\begingroup\@sanitize@url \@url }%
\providecommand \@url [1]{\endgroup\@href {#1}{\urlprefix }}%
\providecommand \urlprefix  [0]{URL }%
\providecommand \Eprint [0]{\href }%
\providecommand \doibase [0]{http://dx.doi.org/}%
\providecommand \selectlanguage [0]{\@gobble}%
\providecommand \bibinfo  [0]{\@secondoftwo}%
\providecommand \bibfield  [0]{\@secondoftwo}%
\providecommand \translation [1]{[#1]}%
\providecommand \BibitemOpen [0]{}%
\providecommand \bibitemStop [0]{}%
\providecommand \bibitemNoStop [0]{.\EOS\space}%
\providecommand \EOS [0]{\spacefactor3000\relax}%
\providecommand \BibitemShut  [1]{\csname bibitem#1\endcsname}%
\let\auto@bib@innerbib\@empty
\bibitem [{\citenamefont {Bernien}\ \emph {et~al.}(2017)\citenamefont {Bernien}, \citenamefont {Schwartz}, \citenamefont {Keesling}, \citenamefont {Levine}, \citenamefont {Omran}, \citenamefont {Pichler}, \citenamefont {Choi}, \citenamefont {Zibrov}, \citenamefont {Endres}, \citenamefont {Greiner} \emph {et~al.}}]{bernien2017probing}%
  \BibitemOpen
  \bibfield  {author} {\bibinfo {author} {\bibfnamefont {H.}~\bibnamefont {Bernien}}, \bibinfo {author} {\bibfnamefont {S.}~\bibnamefont {Schwartz}}, \bibinfo {author} {\bibfnamefont {A.}~\bibnamefont {Keesling}}, \bibinfo {author} {\bibfnamefont {H.}~\bibnamefont {Levine}}, \bibinfo {author} {\bibfnamefont {A.}~\bibnamefont {Omran}}, \bibinfo {author} {\bibfnamefont {H.}~\bibnamefont {Pichler}}, \bibinfo {author} {\bibfnamefont {S.}~\bibnamefont {Choi}}, \bibinfo {author} {\bibfnamefont {A.~S.}\ \bibnamefont {Zibrov}}, \bibinfo {author} {\bibfnamefont {M.}~\bibnamefont {Endres}}, \bibinfo {author} {\bibfnamefont {M.}~\bibnamefont {Greiner}},  \emph {et~al.},\ }\bibfield  {title} {\enquote {\bibinfo {title} {Probing many-body dynamics on a 51-atom quantum simulator},}\ }\href@noop {} {\bibfield  {journal} {\bibinfo  {journal} {Nature}\ }\textbf {\bibinfo {volume} {551}},\ \bibinfo {pages} {579--584} (\bibinfo {year} {2017})}\BibitemShut {NoStop}%
\bibitem [{\citenamefont {Shtanko}\ \emph {et~al.}(2025)\citenamefont {Shtanko}, \citenamefont {Wang}, \citenamefont {Zhang}, \citenamefont {Harle}, \citenamefont {Seif}, \citenamefont {Movassagh},\ and\ \citenamefont {Minev}}]{shtanko2025uncovering}%
  \BibitemOpen
  \bibfield  {author} {\bibinfo {author} {\bibfnamefont {O.}~\bibnamefont {Shtanko}}, \bibinfo {author} {\bibfnamefont {D.~S.}\ \bibnamefont {Wang}}, \bibinfo {author} {\bibfnamefont {H.}~\bibnamefont {Zhang}}, \bibinfo {author} {\bibfnamefont {N.}~\bibnamefont {Harle}}, \bibinfo {author} {\bibfnamefont {A.}~\bibnamefont {Seif}}, \bibinfo {author} {\bibfnamefont {R.}~\bibnamefont {Movassagh}}, \ and\ \bibinfo {author} {\bibfnamefont {Z.}~\bibnamefont {Minev}},\ }\bibfield  {title} {\enquote {\bibinfo {title} {Uncovering local integrability in quantum many-body dynamics},}\ }\href@noop {} {\bibfield  {journal} {\bibinfo  {journal} {Nature Communications}\ }\textbf {\bibinfo {volume} {16}},\ \bibinfo {pages} {2552} (\bibinfo {year} {2025})}\BibitemShut {NoStop}%
\bibitem [{\citenamefont {Kim}\ \emph {et~al.}(2023)\citenamefont {Kim}, \citenamefont {Eddins}, \citenamefont {Anand}, \citenamefont {Wei}, \citenamefont {Van Den~Berg}, \citenamefont {Rosenblatt}, \citenamefont {Nayfeh}, \citenamefont {Wu}, \citenamefont {Zaletel}, \citenamefont {Temme} \emph {et~al.}}]{kim2023evidence}%
  \BibitemOpen
  \bibfield  {author} {\bibinfo {author} {\bibfnamefont {Y.}~\bibnamefont {Kim}}, \bibinfo {author} {\bibfnamefont {A.}~\bibnamefont {Eddins}}, \bibinfo {author} {\bibfnamefont {S.}~\bibnamefont {Anand}}, \bibinfo {author} {\bibfnamefont {K.~X.}\ \bibnamefont {Wei}}, \bibinfo {author} {\bibfnamefont {E.}~\bibnamefont {Van Den~Berg}}, \bibinfo {author} {\bibfnamefont {S.}~\bibnamefont {Rosenblatt}}, \bibinfo {author} {\bibfnamefont {H.}~\bibnamefont {Nayfeh}}, \bibinfo {author} {\bibfnamefont {Y.}~\bibnamefont {Wu}}, \bibinfo {author} {\bibfnamefont {M.}~\bibnamefont {Zaletel}}, \bibinfo {author} {\bibfnamefont {K.}~\bibnamefont {Temme}},  \emph {et~al.},\ }\bibfield  {title} {\enquote {\bibinfo {title} {Evidence for the utility of quantum computing before fault tolerance},}\ }\href@noop {} {\bibfield  {journal} {\bibinfo  {journal} {Nature}\ }\textbf {\bibinfo {volume} {618}},\ \bibinfo {pages} {500--505} (\bibinfo {year} {2023})}\BibitemShut {NoStop}%
\bibitem [{\citenamefont {Frey}\ and\ \citenamefont {Rachel}(2022{\natexlab{a}})}]{doi:10.1126/sciadv.abm7652}%
  \BibitemOpen
  \bibfield  {author} {\bibinfo {author} {\bibfnamefont {P.}~\bibnamefont {Frey}}\ and\ \bibinfo {author} {\bibfnamefont {S.}~\bibnamefont {Rachel}},\ }\bibfield  {title} {\enquote {\bibinfo {title} {Realization of a discrete time crystal on 57 qubits of a quantum computer},}\ }\href {\doibase 10.1126/sciadv.abm7652} {\bibfield  {journal} {\bibinfo  {journal} {Science Advances}\ }\textbf {\bibinfo {volume} {8}},\ \bibinfo {pages} {eabm7652} (\bibinfo {year} {2022}{\natexlab{a}})},\ \Eprint {http://arxiv.org/abs/https://www.science.org/doi/pdf/10.1126/sciadv.abm7652} {https://www.science.org/doi/pdf/10.1126/sciadv.abm7652} \BibitemShut {NoStop}%
\bibitem [{\citenamefont {Mi}\ \emph {et~al.}(2022{\natexlab{a}})\citenamefont {Mi}, \citenamefont {Ippoliti}, \citenamefont {Quintana}, \citenamefont {Greene}, \citenamefont {Chen}, \citenamefont {Gross}, \citenamefont {Arute}, \citenamefont {Arya}, \citenamefont {Atalaya}, \citenamefont {Babbush} \emph {et~al.}}]{mi2022time}%
  \BibitemOpen
  \bibfield  {author} {\bibinfo {author} {\bibfnamefont {X.}~\bibnamefont {Mi}}, \bibinfo {author} {\bibfnamefont {M.}~\bibnamefont {Ippoliti}}, \bibinfo {author} {\bibfnamefont {C.}~\bibnamefont {Quintana}}, \bibinfo {author} {\bibfnamefont {A.}~\bibnamefont {Greene}}, \bibinfo {author} {\bibfnamefont {Z.}~\bibnamefont {Chen}}, \bibinfo {author} {\bibfnamefont {J.}~\bibnamefont {Gross}}, \bibinfo {author} {\bibfnamefont {F.}~\bibnamefont {Arute}}, \bibinfo {author} {\bibfnamefont {K.}~\bibnamefont {Arya}}, \bibinfo {author} {\bibfnamefont {J.}~\bibnamefont {Atalaya}}, \bibinfo {author} {\bibfnamefont {R.}~\bibnamefont {Babbush}},  \emph {et~al.},\ }\bibfield  {title} {\enquote {\bibinfo {title} {Time-crystalline eigenstate order on a quantum processor},}\ }\href@noop {} {\bibfield  {journal} {\bibinfo  {journal} {Nature}\ }\textbf {\bibinfo {volume} {601}},\ \bibinfo {pages} {531--536} (\bibinfo {year} {2022}{\natexlab{a}})}\BibitemShut {NoStop}%
\bibitem [{\citenamefont {Mi}\ \emph {et~al.}(2022{\natexlab{b}})\citenamefont {Mi}, \citenamefont {Sonner}, \citenamefont {Niu}, \citenamefont {Lee}, \citenamefont {Foxen}, \citenamefont {Acharya}, \citenamefont {Aleiner}, \citenamefont {Andersen}, \citenamefont {Arute}, \citenamefont {Arya}, \citenamefont {Asfaw}, \citenamefont {Atalaya}, \citenamefont {Bardin}, \citenamefont {Basso}, \citenamefont {Bengtsson}, \citenamefont {Bortoli}, \citenamefont {Bourassa}, \citenamefont {Brill}, \citenamefont {Broughton}, \citenamefont {Buckley}, \citenamefont {Buell}, \citenamefont {Burkett}, \citenamefont {Bushnell}, \citenamefont {Chen}, \citenamefont {Chiaro}, \citenamefont {Collins}, \citenamefont {Conner}, \citenamefont {Courtney}, \citenamefont {Crook}, \citenamefont {Debroy}, \citenamefont {Demura}, \citenamefont {Dunsworth}, \citenamefont {Eppens}, \citenamefont {Erickson}, \citenamefont {Faoro}, \citenamefont {Farhi}, \citenamefont {Fatemi}, \citenamefont {Flores}, \citenamefont {Forati}, \citenamefont
  {Fowler}, \citenamefont {Giang}, \citenamefont {Gidney}, \citenamefont {Gilboa}, \citenamefont {Giustina}, \citenamefont {Dau}, \citenamefont {Gross}, \citenamefont {Habegger}, \citenamefont {Harrigan}, \citenamefont {Hoffmann}, \citenamefont {Hong}, \citenamefont {Huang}, \citenamefont {Huff}, \citenamefont {Huggins}, \citenamefont {Ioffe}, \citenamefont {Isakov}, \citenamefont {Iveland}, \citenamefont {Jeffrey}, \citenamefont {Jiang}, \citenamefont {Jones}, \citenamefont {Kafri}, \citenamefont {Kechedzhi}, \citenamefont {Khattar}, \citenamefont {Kim}, \citenamefont {Kitaev}, \citenamefont {Klimov}, \citenamefont {Klots}, \citenamefont {Korotkov}, \citenamefont {Kostritsa}, \citenamefont {Kreikebaum}, \citenamefont {Landhuis}, \citenamefont {Laptev}, \citenamefont {Lau}, \citenamefont {Lee}, \citenamefont {Laws}, \citenamefont {Liu}, \citenamefont {Locharla}, \citenamefont {Martin}, \citenamefont {McClean}, \citenamefont {McEwen}, \citenamefont {Costa}, \citenamefont {Miao}, \citenamefont {Mohseni},
  \citenamefont {Montazeri}, \citenamefont {Morvan}, \citenamefont {Mount}, \citenamefont {Mruczkiewicz}, \citenamefont {Naaman}, \citenamefont {Neeley}, \citenamefont {Neill}, \citenamefont {Newman}, \citenamefont {O’Brien}, \citenamefont {Opremcak}, \citenamefont {Petukhov}, \citenamefont {Potter}, \citenamefont {Quintana}, \citenamefont {Rubin}, \citenamefont {Saei}, \citenamefont {Sank}, \citenamefont {Sankaragomathi}, \citenamefont {Satzinger}, \citenamefont {Schuster}, \citenamefont {Shearn}, \citenamefont {Shvarts}, \citenamefont {Strain}, \citenamefont {Su}, \citenamefont {Szalay}, \citenamefont {Vidal}, \citenamefont {Villalonga}, \citenamefont {Vollgraff-Heidweiller}, \citenamefont {White}, \citenamefont {Yao}, \citenamefont {Yeh}, \citenamefont {Yoo}, \citenamefont {Zalcman}, \citenamefont {Zhang}, \citenamefont {Zhu}, \citenamefont {Neven}, \citenamefont {Bacon}, \citenamefont {Hilton}, \citenamefont {Lucero}, \citenamefont {Babbush}, \citenamefont {Boixo}, \citenamefont {Megrant}, \citenamefont
  {Chen}, \citenamefont {Kelly}, \citenamefont {Smelyanskiy}, \citenamefont {Abanin},\ and\ \citenamefont {Roushan}}]{doi:10.1126/science.abq5769}%
  \BibitemOpen
  \bibfield  {author} {\bibinfo {author} {\bibfnamefont {X.}~\bibnamefont {Mi}}, \bibinfo {author} {\bibfnamefont {M.}~\bibnamefont {Sonner}}, \bibinfo {author} {\bibfnamefont {M.~Y.}\ \bibnamefont {Niu}}, \bibinfo {author} {\bibfnamefont {K.~W.}\ \bibnamefont {Lee}}, \bibinfo {author} {\bibfnamefont {B.}~\bibnamefont {Foxen}}, \bibinfo {author} {\bibfnamefont {R.}~\bibnamefont {Acharya}}, \bibinfo {author} {\bibfnamefont {I.}~\bibnamefont {Aleiner}}, \bibinfo {author} {\bibfnamefont {T.~I.}\ \bibnamefont {Andersen}}, \bibinfo {author} {\bibfnamefont {F.}~\bibnamefont {Arute}}, \bibinfo {author} {\bibfnamefont {K.}~\bibnamefont {Arya}}, \bibinfo {author} {\bibfnamefont {A.}~\bibnamefont {Asfaw}}, \bibinfo {author} {\bibfnamefont {J.}~\bibnamefont {Atalaya}}, \bibinfo {author} {\bibfnamefont {J.~C.}\ \bibnamefont {Bardin}}, \bibinfo {author} {\bibfnamefont {J.}~\bibnamefont {Basso}}, \bibinfo {author} {\bibfnamefont {A.}~\bibnamefont {Bengtsson}}, \bibinfo {author} {\bibfnamefont {G.}~\bibnamefont {Bortoli}},
  \bibinfo {author} {\bibfnamefont {A.}~\bibnamefont {Bourassa}}, \bibinfo {author} {\bibfnamefont {L.}~\bibnamefont {Brill}}, \bibinfo {author} {\bibfnamefont {M.}~\bibnamefont {Broughton}}, \bibinfo {author} {\bibfnamefont {B.~B.}\ \bibnamefont {Buckley}}, \bibinfo {author} {\bibfnamefont {D.~A.}\ \bibnamefont {Buell}}, \bibinfo {author} {\bibfnamefont {B.}~\bibnamefont {Burkett}}, \bibinfo {author} {\bibfnamefont {N.}~\bibnamefont {Bushnell}}, \bibinfo {author} {\bibfnamefont {Z.}~\bibnamefont {Chen}}, \bibinfo {author} {\bibfnamefont {B.}~\bibnamefont {Chiaro}}, \bibinfo {author} {\bibfnamefont {R.}~\bibnamefont {Collins}}, \bibinfo {author} {\bibfnamefont {P.}~\bibnamefont {Conner}}, \bibinfo {author} {\bibfnamefont {W.}~\bibnamefont {Courtney}}, \bibinfo {author} {\bibfnamefont {A.~L.}\ \bibnamefont {Crook}}, \bibinfo {author} {\bibfnamefont {D.~M.}\ \bibnamefont {Debroy}}, \bibinfo {author} {\bibfnamefont {S.}~\bibnamefont {Demura}}, \bibinfo {author} {\bibfnamefont {A.}~\bibnamefont {Dunsworth}},
  \bibinfo {author} {\bibfnamefont {D.}~\bibnamefont {Eppens}}, \bibinfo {author} {\bibfnamefont {C.}~\bibnamefont {Erickson}}, \bibinfo {author} {\bibfnamefont {L.}~\bibnamefont {Faoro}}, \bibinfo {author} {\bibfnamefont {E.}~\bibnamefont {Farhi}}, \bibinfo {author} {\bibfnamefont {R.}~\bibnamefont {Fatemi}}, \bibinfo {author} {\bibfnamefont {L.}~\bibnamefont {Flores}}, \bibinfo {author} {\bibfnamefont {E.}~\bibnamefont {Forati}}, \bibinfo {author} {\bibfnamefont {A.~G.}\ \bibnamefont {Fowler}}, \bibinfo {author} {\bibfnamefont {W.}~\bibnamefont {Giang}}, \bibinfo {author} {\bibfnamefont {C.}~\bibnamefont {Gidney}}, \bibinfo {author} {\bibfnamefont {D.}~\bibnamefont {Gilboa}}, \bibinfo {author} {\bibfnamefont {M.}~\bibnamefont {Giustina}}, \bibinfo {author} {\bibfnamefont {A.~G.}\ \bibnamefont {Dau}}, \bibinfo {author} {\bibfnamefont {J.~A.}\ \bibnamefont {Gross}}, \bibinfo {author} {\bibfnamefont {S.}~\bibnamefont {Habegger}}, \bibinfo {author} {\bibfnamefont {M.~P.}\ \bibnamefont {Harrigan}}, \bibinfo
  {author} {\bibfnamefont {M.}~\bibnamefont {Hoffmann}}, \bibinfo {author} {\bibfnamefont {S.}~\bibnamefont {Hong}}, \bibinfo {author} {\bibfnamefont {T.}~\bibnamefont {Huang}}, \bibinfo {author} {\bibfnamefont {A.}~\bibnamefont {Huff}}, \bibinfo {author} {\bibfnamefont {W.~J.}\ \bibnamefont {Huggins}}, \bibinfo {author} {\bibfnamefont {L.~B.}\ \bibnamefont {Ioffe}}, \bibinfo {author} {\bibfnamefont {S.~V.}\ \bibnamefont {Isakov}}, \bibinfo {author} {\bibfnamefont {J.}~\bibnamefont {Iveland}}, \bibinfo {author} {\bibfnamefont {E.}~\bibnamefont {Jeffrey}}, \bibinfo {author} {\bibfnamefont {Z.}~\bibnamefont {Jiang}}, \bibinfo {author} {\bibfnamefont {C.}~\bibnamefont {Jones}}, \bibinfo {author} {\bibfnamefont {D.}~\bibnamefont {Kafri}}, \bibinfo {author} {\bibfnamefont {K.}~\bibnamefont {Kechedzhi}}, \bibinfo {author} {\bibfnamefont {T.}~\bibnamefont {Khattar}}, \bibinfo {author} {\bibfnamefont {S.}~\bibnamefont {Kim}}, \bibinfo {author} {\bibfnamefont {A.~Y.}\ \bibnamefont {Kitaev}}, \bibinfo {author}
  {\bibfnamefont {P.~V.}\ \bibnamefont {Klimov}}, \bibinfo {author} {\bibfnamefont {A.~R.}\ \bibnamefont {Klots}}, \bibinfo {author} {\bibfnamefont {A.~N.}\ \bibnamefont {Korotkov}}, \bibinfo {author} {\bibfnamefont {F.}~\bibnamefont {Kostritsa}}, \bibinfo {author} {\bibfnamefont {J.~M.}\ \bibnamefont {Kreikebaum}}, \bibinfo {author} {\bibfnamefont {D.}~\bibnamefont {Landhuis}}, \bibinfo {author} {\bibfnamefont {P.}~\bibnamefont {Laptev}}, \bibinfo {author} {\bibfnamefont {K.-M.}\ \bibnamefont {Lau}}, \bibinfo {author} {\bibfnamefont {J.}~\bibnamefont {Lee}}, \bibinfo {author} {\bibfnamefont {L.}~\bibnamefont {Laws}}, \bibinfo {author} {\bibfnamefont {W.}~\bibnamefont {Liu}}, \bibinfo {author} {\bibfnamefont {A.}~\bibnamefont {Locharla}}, \bibinfo {author} {\bibfnamefont {O.}~\bibnamefont {Martin}}, \bibinfo {author} {\bibfnamefont {J.~R.}\ \bibnamefont {McClean}}, \bibinfo {author} {\bibfnamefont {M.}~\bibnamefont {McEwen}}, \bibinfo {author} {\bibfnamefont {B.~M.}\ \bibnamefont {Costa}}, \bibinfo {author}
  {\bibfnamefont {K.~C.}\ \bibnamefont {Miao}}, \bibinfo {author} {\bibfnamefont {M.}~\bibnamefont {Mohseni}}, \bibinfo {author} {\bibfnamefont {S.}~\bibnamefont {Montazeri}}, \bibinfo {author} {\bibfnamefont {A.}~\bibnamefont {Morvan}}, \bibinfo {author} {\bibfnamefont {E.}~\bibnamefont {Mount}}, \bibinfo {author} {\bibfnamefont {W.}~\bibnamefont {Mruczkiewicz}}, \bibinfo {author} {\bibfnamefont {O.}~\bibnamefont {Naaman}}, \bibinfo {author} {\bibfnamefont {M.}~\bibnamefont {Neeley}}, \bibinfo {author} {\bibfnamefont {C.}~\bibnamefont {Neill}}, \bibinfo {author} {\bibfnamefont {M.}~\bibnamefont {Newman}}, \bibinfo {author} {\bibfnamefont {T.~E.}\ \bibnamefont {O’Brien}}, \bibinfo {author} {\bibfnamefont {A.}~\bibnamefont {Opremcak}}, \bibinfo {author} {\bibfnamefont {A.}~\bibnamefont {Petukhov}}, \bibinfo {author} {\bibfnamefont {R.}~\bibnamefont {Potter}}, \bibinfo {author} {\bibfnamefont {C.}~\bibnamefont {Quintana}}, \bibinfo {author} {\bibfnamefont {N.~C.}\ \bibnamefont {Rubin}}, \bibinfo {author}
  {\bibfnamefont {N.}~\bibnamefont {Saei}}, \bibinfo {author} {\bibfnamefont {D.}~\bibnamefont {Sank}}, \bibinfo {author} {\bibfnamefont {K.}~\bibnamefont {Sankaragomathi}}, \bibinfo {author} {\bibfnamefont {K.~J.}\ \bibnamefont {Satzinger}}, \bibinfo {author} {\bibfnamefont {C.}~\bibnamefont {Schuster}}, \bibinfo {author} {\bibfnamefont {M.~J.}\ \bibnamefont {Shearn}}, \bibinfo {author} {\bibfnamefont {V.}~\bibnamefont {Shvarts}}, \bibinfo {author} {\bibfnamefont {D.}~\bibnamefont {Strain}}, \bibinfo {author} {\bibfnamefont {Y.}~\bibnamefont {Su}}, \bibinfo {author} {\bibfnamefont {M.}~\bibnamefont {Szalay}}, \bibinfo {author} {\bibfnamefont {G.}~\bibnamefont {Vidal}}, \bibinfo {author} {\bibfnamefont {B.}~\bibnamefont {Villalonga}}, \bibinfo {author} {\bibfnamefont {C.}~\bibnamefont {Vollgraff-Heidweiller}}, \bibinfo {author} {\bibfnamefont {T.}~\bibnamefont {White}}, \bibinfo {author} {\bibfnamefont {Z.}~\bibnamefont {Yao}}, \bibinfo {author} {\bibfnamefont {P.}~\bibnamefont {Yeh}}, \bibinfo {author}
  {\bibfnamefont {J.}~\bibnamefont {Yoo}}, \bibinfo {author} {\bibfnamefont {A.}~\bibnamefont {Zalcman}}, \bibinfo {author} {\bibfnamefont {Y.}~\bibnamefont {Zhang}}, \bibinfo {author} {\bibfnamefont {N.}~\bibnamefont {Zhu}}, \bibinfo {author} {\bibfnamefont {H.}~\bibnamefont {Neven}}, \bibinfo {author} {\bibfnamefont {D.}~\bibnamefont {Bacon}}, \bibinfo {author} {\bibfnamefont {J.}~\bibnamefont {Hilton}}, \bibinfo {author} {\bibfnamefont {E.}~\bibnamefont {Lucero}}, \bibinfo {author} {\bibfnamefont {R.}~\bibnamefont {Babbush}}, \bibinfo {author} {\bibfnamefont {S.}~\bibnamefont {Boixo}}, \bibinfo {author} {\bibfnamefont {A.}~\bibnamefont {Megrant}}, \bibinfo {author} {\bibfnamefont {Y.}~\bibnamefont {Chen}}, \bibinfo {author} {\bibfnamefont {J.}~\bibnamefont {Kelly}}, \bibinfo {author} {\bibfnamefont {V.}~\bibnamefont {Smelyanskiy}}, \bibinfo {author} {\bibfnamefont {D.~A.}\ \bibnamefont {Abanin}}, \ and\ \bibinfo {author} {\bibfnamefont {P.}~\bibnamefont {Roushan}},\ }\bibfield  {title} {\enquote {\bibinfo
  {title} {Noise-resilient edge modes on a chain of superconducting qubits},}\ }\href {\doibase 10.1126/science.abq5769} {\bibfield  {journal} {\bibinfo  {journal} {Science}\ }\textbf {\bibinfo {volume} {378}},\ \bibinfo {pages} {785--790} (\bibinfo {year} {2022}{\natexlab{b}})},\ \Eprint {http://arxiv.org/abs/https://www.science.org/doi/pdf/10.1126/science.abq5769} {https://www.science.org/doi/pdf/10.1126/science.abq5769} \BibitemShut {NoStop}%
\bibitem [{\citenamefont {He}\ \emph {et~al.}(2025)\citenamefont {He}, \citenamefont {Ye}, \citenamefont {Gong}, \citenamefont {Yao}, \citenamefont {Liu}, \citenamefont {Murch}, \citenamefont {Yao},\ and\ \citenamefont {Zu}}]{PhysRevX.15.011055}%
  \BibitemOpen
  \bibfield  {author} {\bibinfo {author} {\bibfnamefont {G.}~\bibnamefont {He}}, \bibinfo {author} {\bibfnamefont {B.}~\bibnamefont {Ye}}, \bibinfo {author} {\bibfnamefont {R.}~\bibnamefont {Gong}}, \bibinfo {author} {\bibfnamefont {C.}~\bibnamefont {Yao}}, \bibinfo {author} {\bibfnamefont {Z.}~\bibnamefont {Liu}}, \bibinfo {author} {\bibfnamefont {K.~W.}\ \bibnamefont {Murch}}, \bibinfo {author} {\bibfnamefont {N.~Y.}\ \bibnamefont {Yao}}, \ and\ \bibinfo {author} {\bibfnamefont {C.}~\bibnamefont {Zu}},\ }\bibfield  {title} {\enquote {\bibinfo {title} {Experimental realization of discrete time quasicrystals},}\ }\href {\doibase 10.1103/PhysRevX.15.011055} {\bibfield  {journal} {\bibinfo  {journal} {Phys. Rev. X}\ }\textbf {\bibinfo {volume} {15}},\ \bibinfo {pages} {011055} (\bibinfo {year} {2025})}\BibitemShut {NoStop}%
\bibitem [{\citenamefont {Shinjo}\ \emph {et~al.}(2024)\citenamefont {Shinjo}, \citenamefont {Seki}, \citenamefont {Shirakawa}, \citenamefont {Sun},\ and\ \citenamefont {Yunoki}}]{shinjo2024unveiling}%
  \BibitemOpen
  \bibfield  {author} {\bibinfo {author} {\bibfnamefont {K.}~\bibnamefont {Shinjo}}, \bibinfo {author} {\bibfnamefont {K.}~\bibnamefont {Seki}}, \bibinfo {author} {\bibfnamefont {T.}~\bibnamefont {Shirakawa}}, \bibinfo {author} {\bibfnamefont {R.-Y.}\ \bibnamefont {Sun}}, \ and\ \bibinfo {author} {\bibfnamefont {S.}~\bibnamefont {Yunoki}},\ }\bibfield  {title} {\enquote {\bibinfo {title} {Unveiling clean two-dimensional discrete time quasicrystals on a digital quantum computer},}\ }\href@noop {} {\bibfield  {journal} {\bibinfo  {journal} {arXiv preprint arXiv:2403.16718}\ } (\bibinfo {year} {2024})}\BibitemShut {NoStop}%
\bibitem [{\citenamefont {Semeghini}\ \emph {et~al.}(2021)\citenamefont {Semeghini}, \citenamefont {Levine}, \citenamefont {Keesling}, \citenamefont {Ebadi}, \citenamefont {Wang}, \citenamefont {Bluvstein}, \citenamefont {Verresen}, \citenamefont {Pichler}, \citenamefont {Kalinowski}, \citenamefont {Samajdar}, \citenamefont {Omran}, \citenamefont {Sachdev}, \citenamefont {Vishwanath}, \citenamefont {Greiner}, \citenamefont {Vuleti^^c4^^87},\ and\ \citenamefont {Lukin}}]{doi:10.1126/science.abi8794}%
  \BibitemOpen
  \bibfield  {author} {\bibinfo {author} {\bibfnamefont {G.}~\bibnamefont {Semeghini}}, \bibinfo {author} {\bibfnamefont {H.}~\bibnamefont {Levine}}, \bibinfo {author} {\bibfnamefont {A.}~\bibnamefont {Keesling}}, \bibinfo {author} {\bibfnamefont {S.}~\bibnamefont {Ebadi}}, \bibinfo {author} {\bibfnamefont {T.~T.}\ \bibnamefont {Wang}}, \bibinfo {author} {\bibfnamefont {D.}~\bibnamefont {Bluvstein}}, \bibinfo {author} {\bibfnamefont {R.}~\bibnamefont {Verresen}}, \bibinfo {author} {\bibfnamefont {H.}~\bibnamefont {Pichler}}, \bibinfo {author} {\bibfnamefont {M.}~\bibnamefont {Kalinowski}}, \bibinfo {author} {\bibfnamefont {R.}~\bibnamefont {Samajdar}}, \bibinfo {author} {\bibfnamefont {A.}~\bibnamefont {Omran}}, \bibinfo {author} {\bibfnamefont {S.}~\bibnamefont {Sachdev}}, \bibinfo {author} {\bibfnamefont {A.}~\bibnamefont {Vishwanath}}, \bibinfo {author} {\bibfnamefont {M.}~\bibnamefont {Greiner}}, \bibinfo {author} {\bibfnamefont {V.}~\bibnamefont {Vuleti^^c4^^87}}, \ and\ \bibinfo {author} {\bibfnamefont
  {M.~D.}\ \bibnamefont {Lukin}},\ }\bibfield  {title} {\enquote {\bibinfo {title} {Probing topological spin liquids on a programmable quantum simulator},}\ }\href {\doibase 10.1126/science.abi8794} {\bibfield  {journal} {\bibinfo  {journal} {Science}\ }\textbf {\bibinfo {volume} {374}},\ \bibinfo {pages} {1242--1247} (\bibinfo {year} {2021})},\ \Eprint {http://arxiv.org/abs/https://www.science.org/doi/pdf/10.1126/science.abi8794} {https://www.science.org/doi/pdf/10.1126/science.abi8794} \BibitemShut {NoStop}%
\bibitem [{\citenamefont {Cai}\ \emph {et~al.}(2023)\citenamefont {Cai}, \citenamefont {Babbush}, \citenamefont {Benjamin}, \citenamefont {Endo}, \citenamefont {Huggins}, \citenamefont {Li}, \citenamefont {McClean},\ and\ \citenamefont {O'Brien}}]{RevModPhys.95.045005}%
  \BibitemOpen
  \bibfield  {author} {\bibinfo {author} {\bibfnamefont {Z.}~\bibnamefont {Cai}}, \bibinfo {author} {\bibfnamefont {R.}~\bibnamefont {Babbush}}, \bibinfo {author} {\bibfnamefont {S.~C.}\ \bibnamefont {Benjamin}}, \bibinfo {author} {\bibfnamefont {S.}~\bibnamefont {Endo}}, \bibinfo {author} {\bibfnamefont {W.~J.}\ \bibnamefont {Huggins}}, \bibinfo {author} {\bibfnamefont {Y.}~\bibnamefont {Li}}, \bibinfo {author} {\bibfnamefont {J.~R.}\ \bibnamefont {McClean}}, \ and\ \bibinfo {author} {\bibfnamefont {T.~E.}\ \bibnamefont {O'Brien}},\ }\bibfield  {title} {\enquote {\bibinfo {title} {Quantum error mitigation},}\ }\href {\doibase 10.1103/RevModPhys.95.045005} {\bibfield  {journal} {\bibinfo  {journal} {Rev. Mod. Phys.}\ }\textbf {\bibinfo {volume} {95}},\ \bibinfo {pages} {045005} (\bibinfo {year} {2023})}\BibitemShut {NoStop}%
\bibitem [{\citenamefont {Endo}, \citenamefont {Benjamin},\ and\ \citenamefont {Li}(2018)}]{PhysRevX.8.031027}%
  \BibitemOpen
  \bibfield  {author} {\bibinfo {author} {\bibfnamefont {S.}~\bibnamefont {Endo}}, \bibinfo {author} {\bibfnamefont {S.~C.}\ \bibnamefont {Benjamin}}, \ and\ \bibinfo {author} {\bibfnamefont {Y.}~\bibnamefont {Li}},\ }\bibfield  {title} {\enquote {\bibinfo {title} {Practical quantum error mitigation for near-future applications},}\ }\href {\doibase 10.1103/PhysRevX.8.031027} {\bibfield  {journal} {\bibinfo  {journal} {Phys. Rev. X}\ }\textbf {\bibinfo {volume} {8}},\ \bibinfo {pages} {031027} (\bibinfo {year} {2018})}\BibitemShut {NoStop}%
\bibitem [{\citenamefont {Koczor}(2021)}]{PhysRevX.11.031057}%
  \BibitemOpen
  \bibfield  {author} {\bibinfo {author} {\bibfnamefont {B.}~\bibnamefont {Koczor}},\ }\bibfield  {title} {\enquote {\bibinfo {title} {Exponential error suppression for near-term quantum devices},}\ }\href {\doibase 10.1103/PhysRevX.11.031057} {\bibfield  {journal} {\bibinfo  {journal} {Phys. Rev. X}\ }\textbf {\bibinfo {volume} {11}},\ \bibinfo {pages} {031057} (\bibinfo {year} {2021})}\BibitemShut {NoStop}%
\bibitem [{\citenamefont {Li}\ and\ \citenamefont {Benjamin}(2017)}]{PhysRevX.7.021050}%
  \BibitemOpen
  \bibfield  {author} {\bibinfo {author} {\bibfnamefont {Y.}~\bibnamefont {Li}}\ and\ \bibinfo {author} {\bibfnamefont {S.~C.}\ \bibnamefont {Benjamin}},\ }\bibfield  {title} {\enquote {\bibinfo {title} {Efficient variational quantum simulator incorporating active error minimization},}\ }\href {\doibase 10.1103/PhysRevX.7.021050} {\bibfield  {journal} {\bibinfo  {journal} {Phys. Rev. X}\ }\textbf {\bibinfo {volume} {7}},\ \bibinfo {pages} {021050} (\bibinfo {year} {2017})}\BibitemShut {NoStop}%
\bibitem [{\citenamefont {Temme}, \citenamefont {Bravyi},\ and\ \citenamefont {Gambetta}(2017)}]{PhysRevLett.119.180509}%
  \BibitemOpen
  \bibfield  {author} {\bibinfo {author} {\bibfnamefont {K.}~\bibnamefont {Temme}}, \bibinfo {author} {\bibfnamefont {S.}~\bibnamefont {Bravyi}}, \ and\ \bibinfo {author} {\bibfnamefont {J.~M.}\ \bibnamefont {Gambetta}},\ }\bibfield  {title} {\enquote {\bibinfo {title} {Error mitigation for short-depth quantum circuits},}\ }\href {\doibase 10.1103/PhysRevLett.119.180509} {\bibfield  {journal} {\bibinfo  {journal} {Phys. Rev. Lett.}\ }\textbf {\bibinfo {volume} {119}},\ \bibinfo {pages} {180509} (\bibinfo {year} {2017})}\BibitemShut {NoStop}%
\bibitem [{\citenamefont {Zhang}\ \emph {et~al.}(2017)\citenamefont {Zhang}, \citenamefont {Hess}, \citenamefont {Kyprianidis}, \citenamefont {Becker}, \citenamefont {Lee}, \citenamefont {Smith}, \citenamefont {Pagano}, \citenamefont {Potirniche}, \citenamefont {Potter}, \citenamefont {Vishwanath} \emph {et~al.}}]{zhang2017observation}%
  \BibitemOpen
  \bibfield  {author} {\bibinfo {author} {\bibfnamefont {J.}~\bibnamefont {Zhang}}, \bibinfo {author} {\bibfnamefont {P.~W.}\ \bibnamefont {Hess}}, \bibinfo {author} {\bibfnamefont {A.}~\bibnamefont {Kyprianidis}}, \bibinfo {author} {\bibfnamefont {P.}~\bibnamefont {Becker}}, \bibinfo {author} {\bibfnamefont {A.}~\bibnamefont {Lee}}, \bibinfo {author} {\bibfnamefont {J.}~\bibnamefont {Smith}}, \bibinfo {author} {\bibfnamefont {G.}~\bibnamefont {Pagano}}, \bibinfo {author} {\bibfnamefont {I.-D.}\ \bibnamefont {Potirniche}}, \bibinfo {author} {\bibfnamefont {A.~C.}\ \bibnamefont {Potter}}, \bibinfo {author} {\bibfnamefont {A.}~\bibnamefont {Vishwanath}},  \emph {et~al.},\ }\bibfield  {title} {\enquote {\bibinfo {title} {Observation of a discrete time crystal},}\ }\href@noop {} {\bibfield  {journal} {\bibinfo  {journal} {Nature}\ }\textbf {\bibinfo {volume} {543}},\ \bibinfo {pages} {217--220} (\bibinfo {year} {2017})}\BibitemShut {NoStop}%
\bibitem [{\citenamefont {Frey}\ and\ \citenamefont {Rachel}(2022{\natexlab{b}})}]{frey2022realization}%
  \BibitemOpen
  \bibfield  {author} {\bibinfo {author} {\bibfnamefont {P.}~\bibnamefont {Frey}}\ and\ \bibinfo {author} {\bibfnamefont {S.}~\bibnamefont {Rachel}},\ }\bibfield  {title} {\enquote {\bibinfo {title} {Realization of a discrete time crystal on 57 qubits of a quantum computer},}\ }\href@noop {} {\bibfield  {journal} {\bibinfo  {journal} {Science advances}\ }\textbf {\bibinfo {volume} {8}},\ \bibinfo {pages} {eabm7652} (\bibinfo {year} {2022}{\natexlab{b}})}\BibitemShut {NoStop}%
\bibitem [{\citenamefont {Choi}\ \emph {et~al.}(2017)\citenamefont {Choi}, \citenamefont {Choi}, \citenamefont {Landig}, \citenamefont {Kucsko}, \citenamefont {Zhou}, \citenamefont {Isoya}, \citenamefont {Jelezko}, \citenamefont {Onoda}, \citenamefont {Sumiya}, \citenamefont {Khemani} \emph {et~al.}}]{choi2017observation}%
  \BibitemOpen
  \bibfield  {author} {\bibinfo {author} {\bibfnamefont {S.}~\bibnamefont {Choi}}, \bibinfo {author} {\bibfnamefont {J.}~\bibnamefont {Choi}}, \bibinfo {author} {\bibfnamefont {R.}~\bibnamefont {Landig}}, \bibinfo {author} {\bibfnamefont {G.}~\bibnamefont {Kucsko}}, \bibinfo {author} {\bibfnamefont {H.}~\bibnamefont {Zhou}}, \bibinfo {author} {\bibfnamefont {J.}~\bibnamefont {Isoya}}, \bibinfo {author} {\bibfnamefont {F.}~\bibnamefont {Jelezko}}, \bibinfo {author} {\bibfnamefont {S.}~\bibnamefont {Onoda}}, \bibinfo {author} {\bibfnamefont {H.}~\bibnamefont {Sumiya}}, \bibinfo {author} {\bibfnamefont {V.}~\bibnamefont {Khemani}},  \emph {et~al.},\ }\bibfield  {title} {\enquote {\bibinfo {title} {Observation of discrete time-crystalline order in a disordered dipolar many-body system},}\ }\href@noop {} {\bibfield  {journal} {\bibinfo  {journal} {Nature}\ }\textbf {\bibinfo {volume} {543}},\ \bibinfo {pages} {221--225} (\bibinfo {year} {2017})}\BibitemShut {NoStop}%
\bibitem [{\citenamefont {Zaletel}\ \emph {et~al.}(2023)\citenamefont {Zaletel}, \citenamefont {Lukin}, \citenamefont {Monroe}, \citenamefont {Nayak}, \citenamefont {Wilczek},\ and\ \citenamefont {Yao}}]{RevModPhys.95.031001}%
  \BibitemOpen
  \bibfield  {author} {\bibinfo {author} {\bibfnamefont {M.~P.}\ \bibnamefont {Zaletel}}, \bibinfo {author} {\bibfnamefont {M.}~\bibnamefont {Lukin}}, \bibinfo {author} {\bibfnamefont {C.}~\bibnamefont {Monroe}}, \bibinfo {author} {\bibfnamefont {C.}~\bibnamefont {Nayak}}, \bibinfo {author} {\bibfnamefont {F.}~\bibnamefont {Wilczek}}, \ and\ \bibinfo {author} {\bibfnamefont {N.~Y.}\ \bibnamefont {Yao}},\ }\bibfield  {title} {\enquote {\bibinfo {title} {Colloquium: Quantum and classical discrete time crystals},}\ }\href {\doibase 10.1103/RevModPhys.95.031001} {\bibfield  {journal} {\bibinfo  {journal} {Rev. Mod. Phys.}\ }\textbf {\bibinfo {volume} {95}},\ \bibinfo {pages} {031001} (\bibinfo {year} {2023})}\BibitemShut {NoStop}%
\bibitem [{\citenamefont {Watanabe}\ and\ \citenamefont {Oshikawa}(2015)}]{PhysRevLett.114.251603}%
  \BibitemOpen
  \bibfield  {author} {\bibinfo {author} {\bibfnamefont {H.}~\bibnamefont {Watanabe}}\ and\ \bibinfo {author} {\bibfnamefont {M.}~\bibnamefont {Oshikawa}},\ }\bibfield  {title} {\enquote {\bibinfo {title} {Absence of quantum time crystals},}\ }\href {\doibase 10.1103/PhysRevLett.114.251603} {\bibfield  {journal} {\bibinfo  {journal} {Phys. Rev. Lett.}\ }\textbf {\bibinfo {volume} {114}},\ \bibinfo {pages} {251603} (\bibinfo {year} {2015})}\BibitemShut {NoStop}%
\bibitem [{\citenamefont {Yao}\ \emph {et~al.}(2017)\citenamefont {Yao}, \citenamefont {Potter}, \citenamefont {Potirniche},\ and\ \citenamefont {Vishwanath}}]{PhysRevLett.118.030401}%
  \BibitemOpen
  \bibfield  {author} {\bibinfo {author} {\bibfnamefont {N.~Y.}\ \bibnamefont {Yao}}, \bibinfo {author} {\bibfnamefont {A.~C.}\ \bibnamefont {Potter}}, \bibinfo {author} {\bibfnamefont {I.-D.}\ \bibnamefont {Potirniche}}, \ and\ \bibinfo {author} {\bibfnamefont {A.}~\bibnamefont {Vishwanath}},\ }\bibfield  {title} {\enquote {\bibinfo {title} {Discrete time crystals: Rigidity, criticality, and realizations},}\ }\href {\doibase 10.1103/PhysRevLett.118.030401} {\bibfield  {journal} {\bibinfo  {journal} {Phys. Rev. Lett.}\ }\textbf {\bibinfo {volume} {118}},\ \bibinfo {pages} {030401} (\bibinfo {year} {2017})}\BibitemShut {NoStop}%
\bibitem [{\citenamefont {Wallman}\ and\ \citenamefont {Emerson}(2016)}]{PhysRevA.94.052325}%
  \BibitemOpen
  \bibfield  {author} {\bibinfo {author} {\bibfnamefont {J.~J.}\ \bibnamefont {Wallman}}\ and\ \bibinfo {author} {\bibfnamefont {J.}~\bibnamefont {Emerson}},\ }\bibfield  {title} {\enquote {\bibinfo {title} {Noise tailoring for scalable quantum computation via randomized compiling},}\ }\href {\doibase 10.1103/PhysRevA.94.052325} {\bibfield  {journal} {\bibinfo  {journal} {Phys. Rev. A}\ }\textbf {\bibinfo {volume} {94}},\ \bibinfo {pages} {052325} (\bibinfo {year} {2016})}\BibitemShut {NoStop}%
\bibitem [{\citenamefont {Van Den~Berg}\ \emph {et~al.}(2023)\citenamefont {Van Den~Berg}, \citenamefont {Minev}, \citenamefont {Kandala},\ and\ \citenamefont {Temme}}]{van2023probabilistic}%
  \BibitemOpen
  \bibfield  {author} {\bibinfo {author} {\bibfnamefont {E.}~\bibnamefont {Van Den~Berg}}, \bibinfo {author} {\bibfnamefont {Z.~K.}\ \bibnamefont {Minev}}, \bibinfo {author} {\bibfnamefont {A.}~\bibnamefont {Kandala}}, \ and\ \bibinfo {author} {\bibfnamefont {K.}~\bibnamefont {Temme}},\ }\bibfield  {title} {\enquote {\bibinfo {title} {Probabilistic error cancellation with sparse pauli--lindblad models on noisy quantum processors},}\ }\href@noop {} {\bibfield  {journal} {\bibinfo  {journal} {Nature physics}\ }\textbf {\bibinfo {volume} {19}},\ \bibinfo {pages} {1116--1121} (\bibinfo {year} {2023})}\BibitemShut {NoStop}%
\bibitem [{\citenamefont {Lazarides}\ and\ \citenamefont {Moessner}(2017)}]{PhysRevB.95.195135}%
  \BibitemOpen
  \bibfield  {author} {\bibinfo {author} {\bibfnamefont {A.}~\bibnamefont {Lazarides}}\ and\ \bibinfo {author} {\bibfnamefont {R.}~\bibnamefont {Moessner}},\ }\bibfield  {title} {\enquote {\bibinfo {title} {Fate of a discrete time crystal in an open system},}\ }\href {\doibase 10.1103/PhysRevB.95.195135} {\bibfield  {journal} {\bibinfo  {journal} {Phys. Rev. B}\ }\textbf {\bibinfo {volume} {95}},\ \bibinfo {pages} {195135} (\bibinfo {year} {2017})}\BibitemShut {NoStop}%
\bibitem [{\citenamefont {Heyl}, \citenamefont {Pollmann},\ and\ \citenamefont {D\'ora}(2018)}]{PhysRevLett.121.016801}%
  \BibitemOpen
  \bibfield  {author} {\bibinfo {author} {\bibfnamefont {M.}~\bibnamefont {Heyl}}, \bibinfo {author} {\bibfnamefont {F.}~\bibnamefont {Pollmann}}, \ and\ \bibinfo {author} {\bibfnamefont {B.}~\bibnamefont {D\'ora}},\ }\bibfield  {title} {\enquote {\bibinfo {title} {Detecting equilibrium and dynamical quantum phase transitions in ising chains via out-of-time-ordered correlators},}\ }\href {\doibase 10.1103/PhysRevLett.121.016801} {\bibfield  {journal} {\bibinfo  {journal} {Phys. Rev. Lett.}\ }\textbf {\bibinfo {volume} {121}},\ \bibinfo {pages} {016801} (\bibinfo {year} {2018})}\BibitemShut {NoStop}%
\bibitem [{\citenamefont {Bin}\ \emph {et~al.}(2023)\citenamefont {Bin}, \citenamefont {Wan}, \citenamefont {Nori}, \citenamefont {Wu},\ and\ \citenamefont {L\"u}}]{PhysRevB.107.L020202}%
  \BibitemOpen
  \bibfield  {author} {\bibinfo {author} {\bibfnamefont {Q.}~\bibnamefont {Bin}}, \bibinfo {author} {\bibfnamefont {L.-L.}\ \bibnamefont {Wan}}, \bibinfo {author} {\bibfnamefont {F.}~\bibnamefont {Nori}}, \bibinfo {author} {\bibfnamefont {Y.}~\bibnamefont {Wu}}, \ and\ \bibinfo {author} {\bibfnamefont {X.-Y.}\ \bibnamefont {L\"u}},\ }\bibfield  {title} {\enquote {\bibinfo {title} {Out-of-time-order correlation as a witness for topological phase transitions},}\ }\href {\doibase 10.1103/PhysRevB.107.L020202} {\bibfield  {journal} {\bibinfo  {journal} {Phys. Rev. B}\ }\textbf {\bibinfo {volume} {107}},\ \bibinfo {pages} {L020202} (\bibinfo {year} {2023})}\BibitemShut {NoStop}%
\bibitem [{\citenamefont {Kobayashi}\ and\ \citenamefont {Motome}(2025{\natexlab{a}})}]{kobayashi2025quantum}%
  \BibitemOpen
  \bibfield  {author} {\bibinfo {author} {\bibfnamefont {K.}~\bibnamefont {Kobayashi}}\ and\ \bibinfo {author} {\bibfnamefont {Y.}~\bibnamefont {Motome}},\ }\bibfield  {title} {\enquote {\bibinfo {title} {Quantum reservoir probing of quantum phase transitions},}\ }\href@noop {} {\bibfield  {journal} {\bibinfo  {journal} {Nature Communications}\ }\textbf {\bibinfo {volume} {16}},\ \bibinfo {pages} {3871} (\bibinfo {year} {2025}{\natexlab{a}})}\BibitemShut {NoStop}%
\bibitem [{\citenamefont {Kobayashi}\ and\ \citenamefont {Motome}(2025{\natexlab{b}})}]{10.21468/SciPostPhys.18.6.198}%
  \BibitemOpen
  \bibfield  {author} {\bibinfo {author} {\bibfnamefont {K.}~\bibnamefont {Kobayashi}}\ and\ \bibinfo {author} {\bibfnamefont {Y.}~\bibnamefont {Motome}},\ }\bibfield  {title} {\enquote {\bibinfo {title} {{Quantum reservoir probing: An inverse paradigm of quantum reservoir computing for exploring quantum many-body physics}},}\ }\href {\doibase 10.21468/SciPostPhys.18.6.198} {\bibfield  {journal} {\bibinfo  {journal} {SciPost Phys.}\ }\textbf {\bibinfo {volume} {18}},\ \bibinfo {pages} {198} (\bibinfo {year} {2025}{\natexlab{b}})}\BibitemShut {NoStop}%
\bibitem [{\citenamefont {Burnett}\ \emph {et~al.}(2019)\citenamefont {Burnett}, \citenamefont {Bengtsson}, \citenamefont {Scigliuzzo}, \citenamefont {Niepce}, \citenamefont {Kudra}, \citenamefont {Delsing},\ and\ \citenamefont {Bylander}}]{burnett2019decoherence}%
  \BibitemOpen
  \bibfield  {author} {\bibinfo {author} {\bibfnamefont {J.~J.}\ \bibnamefont {Burnett}}, \bibinfo {author} {\bibfnamefont {A.}~\bibnamefont {Bengtsson}}, \bibinfo {author} {\bibfnamefont {M.}~\bibnamefont {Scigliuzzo}}, \bibinfo {author} {\bibfnamefont {D.}~\bibnamefont {Niepce}}, \bibinfo {author} {\bibfnamefont {M.}~\bibnamefont {Kudra}}, \bibinfo {author} {\bibfnamefont {P.}~\bibnamefont {Delsing}}, \ and\ \bibinfo {author} {\bibfnamefont {J.}~\bibnamefont {Bylander}},\ }\bibfield  {title} {\enquote {\bibinfo {title} {Decoherence benchmarking of superconducting qubits},}\ }\href@noop {} {\bibfield  {journal} {\bibinfo  {journal} {npj Quantum Information}\ }\textbf {\bibinfo {volume} {5}},\ \bibinfo {pages} {54} (\bibinfo {year} {2019})}\BibitemShut {NoStop}%
\bibitem [{\citenamefont {Klimov}\ \emph {et~al.}(2018)\citenamefont {Klimov}, \citenamefont {Kelly}, \citenamefont {Chen}, \citenamefont {Neeley}, \citenamefont {Megrant}, \citenamefont {Burkett}, \citenamefont {Barends}, \citenamefont {Arya}, \citenamefont {Chiaro}, \citenamefont {Chen}, \citenamefont {Dunsworth}, \citenamefont {Fowler}, \citenamefont {Foxen}, \citenamefont {Gidney}, \citenamefont {Giustina}, \citenamefont {Graff}, \citenamefont {Huang}, \citenamefont {Jeffrey}, \citenamefont {Lucero}, \citenamefont {Mutus}, \citenamefont {Naaman}, \citenamefont {Neill}, \citenamefont {Quintana}, \citenamefont {Roushan}, \citenamefont {Sank}, \citenamefont {Vainsencher}, \citenamefont {Wenner}, \citenamefont {White}, \citenamefont {Boixo}, \citenamefont {Babbush}, \citenamefont {Smelyanskiy}, \citenamefont {Neven},\ and\ \citenamefont {Martinis}}]{PhysRevLett.121.090502}%
  \BibitemOpen
  \bibfield  {author} {\bibinfo {author} {\bibfnamefont {P.~V.}\ \bibnamefont {Klimov}}, \bibinfo {author} {\bibfnamefont {J.}~\bibnamefont {Kelly}}, \bibinfo {author} {\bibfnamefont {Z.}~\bibnamefont {Chen}}, \bibinfo {author} {\bibfnamefont {M.}~\bibnamefont {Neeley}}, \bibinfo {author} {\bibfnamefont {A.}~\bibnamefont {Megrant}}, \bibinfo {author} {\bibfnamefont {B.}~\bibnamefont {Burkett}}, \bibinfo {author} {\bibfnamefont {R.}~\bibnamefont {Barends}}, \bibinfo {author} {\bibfnamefont {K.}~\bibnamefont {Arya}}, \bibinfo {author} {\bibfnamefont {B.}~\bibnamefont {Chiaro}}, \bibinfo {author} {\bibfnamefont {Y.}~\bibnamefont {Chen}}, \bibinfo {author} {\bibfnamefont {A.}~\bibnamefont {Dunsworth}}, \bibinfo {author} {\bibfnamefont {A.}~\bibnamefont {Fowler}}, \bibinfo {author} {\bibfnamefont {B.}~\bibnamefont {Foxen}}, \bibinfo {author} {\bibfnamefont {C.}~\bibnamefont {Gidney}}, \bibinfo {author} {\bibfnamefont {M.}~\bibnamefont {Giustina}}, \bibinfo {author} {\bibfnamefont {R.}~\bibnamefont {Graff}},
  \bibinfo {author} {\bibfnamefont {T.}~\bibnamefont {Huang}}, \bibinfo {author} {\bibfnamefont {E.}~\bibnamefont {Jeffrey}}, \bibinfo {author} {\bibfnamefont {E.}~\bibnamefont {Lucero}}, \bibinfo {author} {\bibfnamefont {J.~Y.}\ \bibnamefont {Mutus}}, \bibinfo {author} {\bibfnamefont {O.}~\bibnamefont {Naaman}}, \bibinfo {author} {\bibfnamefont {C.}~\bibnamefont {Neill}}, \bibinfo {author} {\bibfnamefont {C.}~\bibnamefont {Quintana}}, \bibinfo {author} {\bibfnamefont {P.}~\bibnamefont {Roushan}}, \bibinfo {author} {\bibfnamefont {D.}~\bibnamefont {Sank}}, \bibinfo {author} {\bibfnamefont {A.}~\bibnamefont {Vainsencher}}, \bibinfo {author} {\bibfnamefont {J.}~\bibnamefont {Wenner}}, \bibinfo {author} {\bibfnamefont {T.~C.}\ \bibnamefont {White}}, \bibinfo {author} {\bibfnamefont {S.}~\bibnamefont {Boixo}}, \bibinfo {author} {\bibfnamefont {R.}~\bibnamefont {Babbush}}, \bibinfo {author} {\bibfnamefont {V.~N.}\ \bibnamefont {Smelyanskiy}}, \bibinfo {author} {\bibfnamefont {H.}~\bibnamefont {Neven}}, \ and\
  \bibinfo {author} {\bibfnamefont {J.~M.}\ \bibnamefont {Martinis}},\ }\bibfield  {title} {\enquote {\bibinfo {title} {Fluctuations of energy-relaxation times in superconducting qubits},}\ }\href {\doibase 10.1103/PhysRevLett.121.090502} {\bibfield  {journal} {\bibinfo  {journal} {Phys. Rev. Lett.}\ }\textbf {\bibinfo {volume} {121}},\ \bibinfo {pages} {090502} (\bibinfo {year} {2018})}\BibitemShut {NoStop}%
\bibitem [{\citenamefont {Schl\"or}\ \emph {et~al.}(2019)\citenamefont {Schl\"or}, \citenamefont {Lisenfeld}, \citenamefont {M\"uller}, \citenamefont {Bilmes}, \citenamefont {Schneider}, \citenamefont {Pappas}, \citenamefont {Ustinov},\ and\ \citenamefont {Weides}}]{PhysRevLett.123.190502}%
  \BibitemOpen
  \bibfield  {author} {\bibinfo {author} {\bibfnamefont {S.}~\bibnamefont {Schl\"or}}, \bibinfo {author} {\bibfnamefont {J.}~\bibnamefont {Lisenfeld}}, \bibinfo {author} {\bibfnamefont {C.}~\bibnamefont {M\"uller}}, \bibinfo {author} {\bibfnamefont {A.}~\bibnamefont {Bilmes}}, \bibinfo {author} {\bibfnamefont {A.}~\bibnamefont {Schneider}}, \bibinfo {author} {\bibfnamefont {D.~P.}\ \bibnamefont {Pappas}}, \bibinfo {author} {\bibfnamefont {A.~V.}\ \bibnamefont {Ustinov}}, \ and\ \bibinfo {author} {\bibfnamefont {M.}~\bibnamefont {Weides}},\ }\bibfield  {title} {\enquote {\bibinfo {title} {Correlating decoherence in transmon qubits: Low frequency noise by single fluctuators},}\ }\href {\doibase 10.1103/PhysRevLett.123.190502} {\bibfield  {journal} {\bibinfo  {journal} {Phys. Rev. Lett.}\ }\textbf {\bibinfo {volume} {123}},\ \bibinfo {pages} {190502} (\bibinfo {year} {2019})}\BibitemShut {NoStop}%
\bibitem [{\citenamefont {Hirasaki}\ \emph {et~al.}(2023)\citenamefont {Hirasaki}, \citenamefont {Daimon}, \citenamefont {Itoko}, \citenamefont {Kanazawa},\ and\ \citenamefont {Saitoh}}]{10.1063/5.0166739}%
  \BibitemOpen
  \bibfield  {author} {\bibinfo {author} {\bibfnamefont {Y.}~\bibnamefont {Hirasaki}}, \bibinfo {author} {\bibfnamefont {S.}~\bibnamefont {Daimon}}, \bibinfo {author} {\bibfnamefont {T.}~\bibnamefont {Itoko}}, \bibinfo {author} {\bibfnamefont {N.}~\bibnamefont {Kanazawa}}, \ and\ \bibinfo {author} {\bibfnamefont {E.}~\bibnamefont {Saitoh}},\ }\bibfield  {title} {\enquote {\bibinfo {title} {Detection of temporal fluctuation in superconducting qubits for quantum error mitigation},}\ }\href {\doibase 10.1063/5.0166739} {\bibfield  {journal} {\bibinfo  {journal} {Applied Physics Letters}\ }\textbf {\bibinfo {volume} {123}},\ \bibinfo {pages} {184002} (\bibinfo {year} {2023})},\ \Eprint {http://arxiv.org/abs/https://pubs.aip.org/aip/apl/article-pdf/doi/10.1063/5.0166739/18193281/184002\_1\_5.0166739.pdf} {https://pubs.aip.org/aip/apl/article-pdf/doi/10.1063/5.0166739/18193281/184002\_1\_5.0166739.pdf} \BibitemShut {NoStop}%
\bibitem [{\citenamefont {Hirasaki}\ \emph {et~al.}(2024)\citenamefont {Hirasaki}, \citenamefont {Daimon}, \citenamefont {Kanazawa}, \citenamefont {Itoko}, \citenamefont {Tokunari},\ and\ \citenamefont {Saitoh}}]{10.1063/5.0226517}%
  \BibitemOpen
  \bibfield  {author} {\bibinfo {author} {\bibfnamefont {Y.}~\bibnamefont {Hirasaki}}, \bibinfo {author} {\bibfnamefont {S.}~\bibnamefont {Daimon}}, \bibinfo {author} {\bibfnamefont {N.}~\bibnamefont {Kanazawa}}, \bibinfo {author} {\bibfnamefont {T.}~\bibnamefont {Itoko}}, \bibinfo {author} {\bibfnamefont {M.}~\bibnamefont {Tokunari}}, \ and\ \bibinfo {author} {\bibfnamefont {E.}~\bibnamefont {Saitoh}},\ }\bibfield  {title} {\enquote {\bibinfo {title} {Dynamics of measurement-induced state transitions in superconducting qubits},}\ }\href {\doibase 10.1063/5.0226517} {\bibfield  {journal} {\bibinfo  {journal} {Journal of Applied Physics}\ }\textbf {\bibinfo {volume} {136}},\ \bibinfo {pages} {124401} (\bibinfo {year} {2024})},\ \Eprint {http://arxiv.org/abs/https://pubs.aip.org/aip/jap/article-pdf/doi/10.1063/5.0226517/20169975/124401\_1\_5.0226517.pdf} {https://pubs.aip.org/aip/jap/article-pdf/doi/10.1063/5.0226517/20169975/124401\_1\_5.0226517.pdf} \BibitemShut {NoStop}%
\bibitem [{\citenamefont {Etxezarreta~Martinez}\ \emph {et~al.}(2021)\citenamefont {Etxezarreta~Martinez}, \citenamefont {Fuentes}, \citenamefont {Crespo},\ and\ \citenamefont {Garcia-Frias}}]{etxezarreta2021time}%
  \BibitemOpen
  \bibfield  {author} {\bibinfo {author} {\bibfnamefont {J.}~\bibnamefont {Etxezarreta~Martinez}}, \bibinfo {author} {\bibfnamefont {P.}~\bibnamefont {Fuentes}}, \bibinfo {author} {\bibfnamefont {P.}~\bibnamefont {Crespo}}, \ and\ \bibinfo {author} {\bibfnamefont {J.}~\bibnamefont {Garcia-Frias}},\ }\bibfield  {title} {\enquote {\bibinfo {title} {Time-varying quantum channel models for superconducting qubits},}\ }\href@noop {} {\bibfield  {journal} {\bibinfo  {journal} {npj Quantum Information}\ }\textbf {\bibinfo {volume} {7}},\ \bibinfo {pages} {115} (\bibinfo {year} {2021})}\BibitemShut {NoStop}%
\bibitem [{\citenamefont {Stehli}\ \emph {et~al.}(2020)\citenamefont {Stehli}, \citenamefont {Brehm}, \citenamefont {Wolz}, \citenamefont {Baity}, \citenamefont {Danilin}, \citenamefont {Seferai}, \citenamefont {Rotzinger}, \citenamefont {Ustinov},\ and\ \citenamefont {Weides}}]{stehli2020coherent}%
  \BibitemOpen
  \bibfield  {author} {\bibinfo {author} {\bibfnamefont {A.}~\bibnamefont {Stehli}}, \bibinfo {author} {\bibfnamefont {J.~D.}\ \bibnamefont {Brehm}}, \bibinfo {author} {\bibfnamefont {T.}~\bibnamefont {Wolz}}, \bibinfo {author} {\bibfnamefont {P.}~\bibnamefont {Baity}}, \bibinfo {author} {\bibfnamefont {S.}~\bibnamefont {Danilin}}, \bibinfo {author} {\bibfnamefont {V.}~\bibnamefont {Seferai}}, \bibinfo {author} {\bibfnamefont {H.}~\bibnamefont {Rotzinger}}, \bibinfo {author} {\bibfnamefont {A.~V.}\ \bibnamefont {Ustinov}}, \ and\ \bibinfo {author} {\bibfnamefont {M.}~\bibnamefont {Weides}},\ }\bibfield  {title} {\enquote {\bibinfo {title} {Coherent superconducting qubits from a subtractive junction fabrication process},}\ }\href@noop {} {\bibfield  {journal} {\bibinfo  {journal} {Applied Physics Letters}\ }\textbf {\bibinfo {volume} {117}} (\bibinfo {year} {2020})}\BibitemShut {NoStop}%
\end{thebibliography}%

\end{document}